\documentclass[twocolumn,showpacs,preprintnumbers,superscriptaddress,amsmath,amssymb, prb]{revtex4}
\usepackage{graphicx}
\usepackage{dcolumn}
\usepackage{bm}

\begin{document}

\title{Evidence for mass renormalization in LaNiO$_3$: an \textit{in situ} soft x-ray photoemission study of epitaxial films}

\author{K.~Horiba}
\email{horiba@spring8.or.jp}
\affiliation {RIKEN SPring-8 Center, Sayo-cho, Sayo-gun, Hyogo 679-5148, Japan}

\author{R.~Eguchi}
\affiliation {RIKEN SPring-8 Center, Sayo-cho, Sayo-gun, Hyogo 679-5148, Japan}

\author{M.~Taguchi}
\affiliation {RIKEN SPring-8 Center, Sayo-cho, Sayo-gun, Hyogo 679-5148, Japan}

\author{A.~Chainani}
\affiliation {RIKEN SPring-8 Center, Sayo-cho, Sayo-gun, Hyogo 679-5148, Japan}

\author{A.~Kikkawa}
\affiliation {RIKEN SPring-8 Center, Sayo-cho, Sayo-gun, Hyogo 679-5148, Japan}

\author{Y.~Senba}
\affiliation {JASRI/SPring-8, Sayo-cho, Sayo-gun, Hyogo 679-5198, Japan}

\author{H.~Ohashi}
\affiliation {JASRI/SPring-8, Sayo-cho, Sayo-gun, Hyogo 679-5198, Japan}

\author{S.~Shin}
\affiliation {RIKEN SPring-8 Center, Sayo-cho, Sayo-gun, Hyogo 679-5148, Japan}
\affiliation {Institute for Solid State Physics, The University of Tokyo, Kashiwa, Chiba 277-8581, Japan}

\date{\today}

\begin{abstract}
We investigate the electronic structure of high-quality single-crystal LaNiO$_3$ (LNO) thin films using {\it in situ} photoemission spectroscopy (PES). The {\it in situ} high-resolution soft x-ray PES measurements on epitaxial thin films reveal the intrinsic electronic structure of LNO. We find a new sharp feature in the PES spectra crossing the Fermi level, which is derived from the correlated Ni 3$d$ $e_g$ electrons. This feature shows significant enhancement of spectral weight with decreasing temperature. From a detailed analysis of resistivity data, the enhancement of spectral weight is attributed to increasing electron correlations due to antiferromagnetic fluctuations.
\end{abstract}


\maketitle

Perovskite-type nickel oxides have attracted considerable attention because of their unusual physical properties such as spin and orbital ordering and metal-insulator transitions derived from the interplay of electronic, structural, and magnetic degrees of freedom. \cite{Goodenough_LNO, Rajeev_LNO, Sreedhar_LNO, Torrance_RNO, Xu_LNO, Zhou_RNO} Among the nickelates of general formula $R$NiO$_3$ ($R$NO; $R$~=~rare earth), LaNiO$_3$ (LNO) is a prototypical compound and exhibits paramagnetic metallic properties at all temperatures. The $T^2$ dependence of resistivity, heat capacity, and susceptibility data suggest an enhanced electron effective mass ($m^*$~$\sim$~10$m_b$), owing to the strongly-correlated Ni 3$d$ $e_g$ electrons \cite{Rajeev_LNO, Sreedhar_LNO, Xu_LNO, Zhou_RNO} close to a metal-insulator transition. Indeed, PrNiO$_3$ and NdNiO$_3$ undergo a temperature-dependent metal to antiferromagnetic (AF)-insulator transition. \cite{Torrance_RNO, Xu_LNO, Zhou_RNO}

In order to address the nature of correlation-induced unusual properties, photoemission spectroscopy (PES) has provided rich information. \cite{Mo_V2O3, Sekiyama_SCVO, Sekiyama_CeRu2, Barman_PES, Mizokawa_RNO, Vobornik_RNO} Recent progress in high-resolution soft x-ray (SX; $h\nu$~$\sim$~1,000~eV) PES has made it possible to reveal the bulk- and 3$d$-sensitive electronic structure of strongly-correlated electron systems. \cite{Mo_V2O3, Sekiyama_SCVO} However, despite the larger photoelectron mean free paths compared to vacuum ultraviolet (VUV; $h\nu$~$\sim$~10~-~100~eV) PES, SX-PES spectra may also strongly depend on the surface condition. \cite{Sekiyama_CeRu2} While PES measurements on thin films are also difficult in terms of surface treatment, very recent work has shown the importance of \textit{in situ} PES on oxide thin films. \cite{Horiba_RSI, Shi_LSMO, Horiba_LSMO} Previous PES studies on the electronic structure of $R$NO have all been carried out on polycrystalline samples and a scraping procedure was used to obtain fresh surfaces. \cite{Barman_PES, Mizokawa_RNO, Vobornik_RNO} Furthermore, while it is extremely difficult to grow single crystals of the $R$NO series, the growth of epitaxial films of $R$NO for use in device applications is known. \cite{Satyalakshmni_LNOfilm, Chen_LNOfilm, Dobin_LNOfilm, Wells_RNOfilm} Hence we felt it important to study the electronic structure of LNO using {\it in situ} SX PES. In this paper, we report \textit{in situ} PES study on high-quality LNO epitaxial thin films, which are also characterized for the structure and by electrical resistivity. Most significantly, we observe a narrow band crossing the Fermi level ($E_{\rm F}$), which shows a temperature-dependent enhancement of spectral weight. From a careful analysis of the resistivity data, we attribute the observed behavior to increase in correlations due to AF fluctuations.  

\begin{figure}
\includegraphics[width=0.94\linewidth]{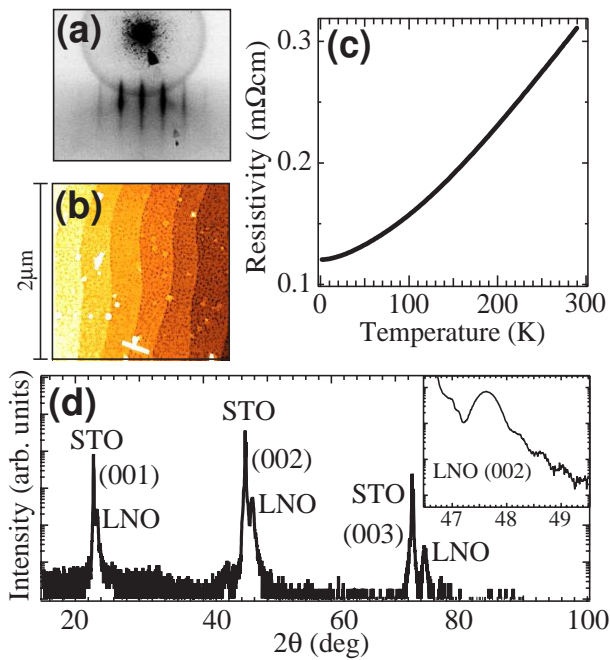}
\caption{\label{figure1} (Color online) (a) RHEED pattern, (b) AFM image, (c) temperature dependence of resistivity, and (d) XRD pattern (Inset: expansion graph in the vicinity of the (002) Bragg peak) of LNO films grown on STO (001) substrates.}
\end{figure}

The fabrication of the LNO thin films and SX spectroscopic measurements were carried out using the high-resolution synchrotron radiation PES system combined with a pulsed laser deposition chamber, which was installed at beamline BL17SU of SPring-8. \cite{Horiba_BL17SESPLD} LNO thin films were grown epitaxially on SrTiO$_3$ (STO) substrates. Sintered stoichiometric LNO pellets were used as ablation targets. A Nd:~YAG laser was used for ablation in its frequency-tripled mode ($\lambda$~=~355~nm) at a repetition rate of 1~Hz. The wet-etched STO (001) substrates were annealed at 900~$^{\circ}$C at an oxygen pressure of 1~$\times$~10$^{-4}$ Pa before deposition. The substrate temperature was set to 650~$^{\circ}$C and the oxygen pressure was 10 Pa during the deposition. The fabricated LNO thin films were subsequently annealed at 400~$^{\circ}$C for 30 minutes in atmospheric pressure of oxygen to remove oxygen vacancies. After cooling the sample to below 100~$^{\circ}$C and evacuating the growth chamber, the surface morphology and crystallinity of the fabricated LNO thin films was checked by \textit{in situ} observation of reflection high-energy electron diffraction (RHEED) pattern. Sharp streak patterns shown in Fig.~\ref{figure1}~(a) indicate the smooth and high-quality single crystal surface of the fabricated LNO thin films.

The PES spectra were obtained using a Gammadata-Scienta SES-2002 electron-energy analyzer. The total energy resolution at the photon energy of 800~eV was set to about 200~meV for typical PES measurements and 70~meV for high-resolution PES measurements. The Fermi level of the samples and the energy resolution of PES spectra were obtained from an evaporated gold film. X-ray absorption (XAS) spectra were obtained by measuring the sample drain current. After PES and XAS measurements, the samples were characterized {\it ex~situ} for the surface morphology by atomic force microscopy (AFM), electrical resistivity by the four-probe method, and crystal structure by x-ray diffraction (XRD) measurements. In the AFM image shown in Fig.~\ref{figure1}~(b), atomically-flat step-and-terrace structures which reflect the surface of STO substrates are clearly observed, indicating that film surfaces can be controlled on an atomic scale. Figure~\ref{figure1}~(c) and (d) show temperature dependence of resistivity and the XRD pattern of fabricated LNO films, respectively. The resistivity values and the position of Bragg peaks are very consistent with the previous reports on LNO thin films. \cite{Satyalakshmni_LNOfilm, Chen_LNOfilm, Dobin_LNOfilm} The fringe pattern around the LNO(002) Bragg peak shown in the inset of Fig.~\ref{figure1}~(d) suggests a thickness uniformity of the order of a unit cell, that is, it confirms the formation of atomically-flat thin film surface and interface between the film and the substrate. \cite{Dobin_LNOfilm} These results indicate that the high-quality pseudomorphic LNO thin films are grown epitaxially on STO (001) substrates. The thickness of the film was estimated to be 20~nm by grazing incidence x-ray reflectivity.

\begin{figure}
\includegraphics[width=0.9\linewidth]{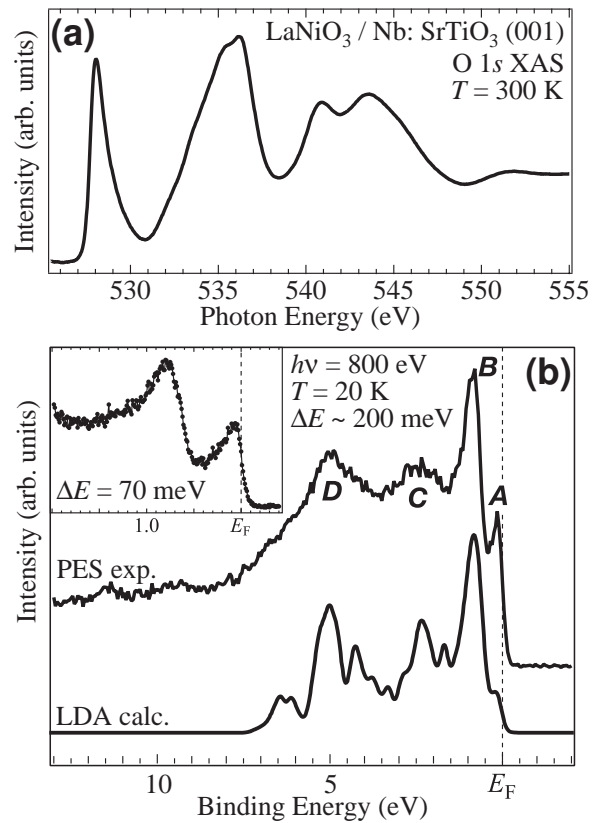}
\caption{\label{figure2} (a) \textit{In situ} O 1$s$ X-ray absorption spectrum  of LNO films grown on Nb-doped STO (001) substrates. (b) \textit{In situ} valence band photoemission spectrum of LNO films and calculated DOS of LNO. The inset shows high-resolution ($\Delta E$~$\sim$~70~meV) photoemission spectrum of near-$E_{\rm F}$ region.}
\end{figure}

{\it In~situ} O 1$s$ XAS and PES spectra of fabricated LNO thin films are shown in Fig.~\ref{figure2}. The O 1$s$ XAS spectrum, which have much deeper probing depth than that of PES measurement, is in excellent agreement with the previous results measured on scraped polycrystalline LNO surfaces. \cite{Medarde_XAS, Abbate_XAS} The structure of the peak around 528~eV, which corresponds to the strongly-hybridized states of O~2$p$ and Ni~3$d$, is known to be very sensitive to the oxygen stoichiometry or the presence of an impulity phase such as layered La$_2$NiO$_4$. \cite{Abbate_XAS} The sharp and strong peak structure of first peak indicates the successful fabrication of stoichiometric single phase LNO thin films.     

In contrast to O 1$s$ XAS spectra, {\it in~situ} PES spectra show an important difference with the previous report on scraped polycrystalline LNO surfaces. \cite{Barman_PES} In the previous study, VUV (He I; $h\nu$~=~21.2~eV) PES showed two weak features near $E_{\rm F}$ (labeled $A$ and $B$, and assigned to $e_g$ and $t_{2g}$ states, respectively) and two prominent structures (labeled $C$ and $D$) which are O~2$p$ dominant states. In our {\it in~situ} PES spectra, the Ni~3$d$ states (labeled $A$ and $B$) are considerably enhanced since the calculated cross section ratio \cite{Yeh_Lindau} between the Ni~3$d$ and O~2$p$ orbitals at the photon energy of 800~eV is 600 times larger than that at 21.2~eV. Barman {\it et~al}. have also reported the PES spectrum using Mg~K$\alpha$ x-ray source, where the cross section ratio is similar to that at 800~eV, but the spectrum shows a single broad peak instead of two features. This difference probably originates in the lower resolution and/or surface quality of the earlier study. Thus, the sharp peak crossing $E_{\rm F}$ is clearly resolved using high-resolution synchrotron radiation {\it in~situ} PES. A higher-resolution spectrum shown in the inset of Fig.~\ref{figure2}~(b) indicates that the peak top of $e_g$ state lies above $E_{\rm F}$.

\begin{figure}
\includegraphics[width=0.85\linewidth]{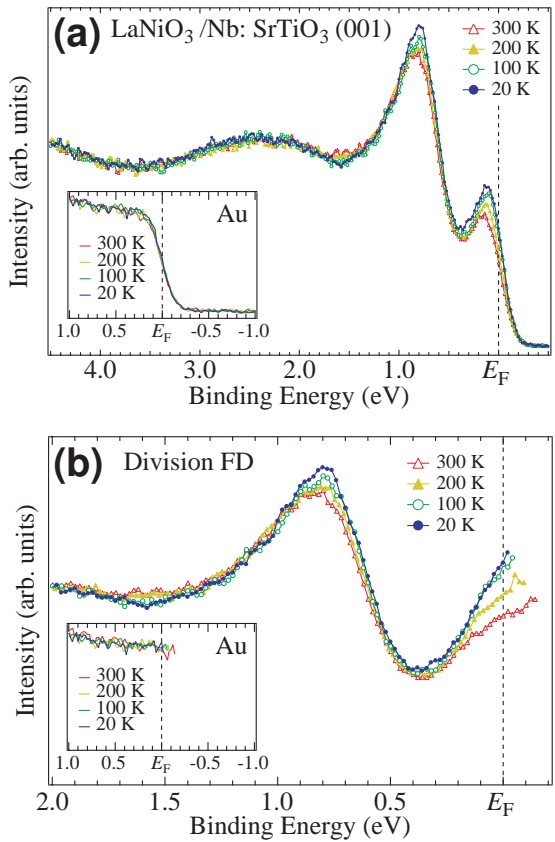}	
\caption{\label{figure3} (Color online) (a) Temperature dependent photoemission spectra of LNO films. (b) Near-$E_{\rm F}$ spectra divided by Fermi-Dirac functions estimated by the gold spectra at each temperature.}
\end{figure}

In Fig.~\ref{figure2}~(b), the calculated density of states (DOS) is plotted with the experimental valence band PES spectrum. The theoretical calculation was done using the local density approximation based full-potential linearized augmented plane-wave method \cite{Wien2k} in the $R\bar{3}c$ symmetry for LNO. For the comparison with the PES spectra, the calculated DOS is multiplied by the Fermi-Dirac function and convoluted by Gaussian function with a width of 0.2 eV. The calculated DOS shows good agreement with the energy position of the four peak structures in the PES spectrum. The agreement with the band calculation suggests that the electronic structure of LNO can be described well within the single-particle description. However, the experimental data indicates higher intensity for the feature crossing $E_{\rm F}$ (labeled $A$). The enhanced spectral intensity indicates a renormalization of electronic states at $E_{\rm F}$, corresponding to the enhanced effective mass known from thermodynamic and transport studies. \cite{Rajeev_LNO, Sreedhar_LNO, Xu_LNO, Zhou_RNO} 

Figure~\ref{figure3} shows the temperature dependence of PES spectra in the near-$E_{\rm F}$ region. On decreasing temperature, the intensity of near-$E_{\rm F}$ features assigned as Ni 3$d$  $t_{2g}$ and $e_g$ states get enhanced. In the spectra divided by Fermi-Dirac function shown in Fig.~\ref{figure3}~(b), we note that the most remarkable change occurs in the feature originating in quarter-filled Ni 3$d$ $e_g$ states crossing $E_{\rm F}$. A similar behavior was observed in another 3$d$ electron system \cite{Shimo_LVO} with large effective electron mass. A Kondo-like enhancement of the $e_g$-derived states was predicted for LNO. \cite{Anisimov}

\begin{figure}
\includegraphics[width=1\linewidth]{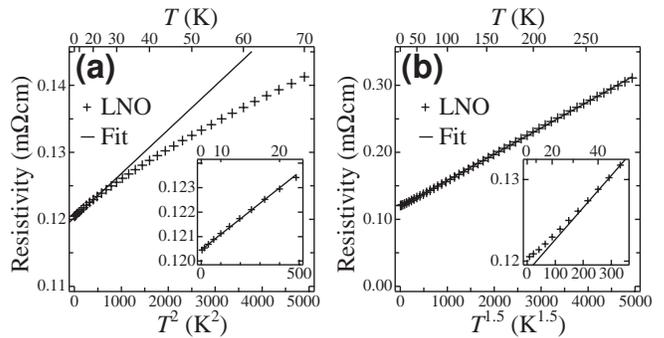}
\caption{\label{figure4} (a) Resistivity vs $T^2$ for LNO thin films below 70 K and a fitted line. The inset shows the expansion graph of Resistivity vs $T^2$ below 20 K and the best fit to $\rho$~=~$\rho_0$~+~$AT^2$; $A$~=~7.1~$\times$~$10^{-6}$~m$\Omega$cm/${\rm K}^2$. (b) Resistivity vs $T^{1.5}$ 
for LNO thin films and the best fit below 300 K to $\rho$~=~$\rho_0'$~+~$A'T^{1.5}$; $A'$~=~3.9~$\times$~$10^{-5}$~m$\Omega$cm/${\rm K}^{1.5}$.  The inset shows the expansion graph at low temperature. }
\end{figure}

In order to explain the origin of the change in the spectra with temperature, we have carried out detailed analyses for the temperature-dependent resistivity data of our LNO films. Figure~\ref{figure4}~(a) and (b) show $T^2$ and $T^{1.5}$ plots of the resistivity data, respectively. According to Landau Fermi-liquid theory, temperature dependence of resistivity follows an equation, $\rho$~=~$\rho_0$~+~$AT^2$, where $\rho_0$ is the residual resistivity and the $T^2$ term is due to electron-electron scattering. At low temperature below 20 K, we can fit our resistivity data well to the preceding equation with the $\rho_0$ and $A$ value of 0.1204~m$\Omega$ and 7.1~$\times$~$10^{-6}$~m$\Omega$cm/${\rm K}^2$, respectively. We assume the electronic specific heat coefficient of our LNO thin films to be same as the value for polycrystalline bulk LNO \cite{Rajeev_LNO, Sreedhar_LNO} $\gamma$~=~13.8~mJ/mol~${\rm K}^2$, because we cannot measure the heat capacity of thin film samples. This implies a Kadowaki-Woods ratio \cite{Kadowaki_Woods} $r_{\rm KW}~\equiv~A/\gamma^2$ for the LNO thin film to be 3.7~$a_0$, where $a_0$~$\equiv$~10~$\mu$$\Omega$cm~${\rm mol}^2$${\rm K}^2$/${\rm J}^2$. This value is close to that for typical correlated $d$ electron system with magnetic frustration or close to a Mott insulator, such as La$_{1.7}$Sr$_{0.3}$CuO$_{4}$, V$_{2}$O$_{3}$, LiV$_{2}$O$_{4}$, and Ca$_{1.8}$Sr$_{0.2}$RuO$_{4}$. \cite{Li_KW} On the other hand, at high temperatures (above 20 K), the resistivity value deviates from $T^2$ and follows a power law, $\rho$~=~$\rho_0'$~+~$A'T^{1.5}$.
The $T^{1.5}$ dependence of resistivity is not a special phenomenon of the LNO thin film and has been reported for bulk LNO polycrystal also. \cite{Xu_LNO} The $T^{1.5}$ dependence of resistivity has been also reported on other correlated systems such as NiS$_{1-x}$Se$_x$ under high pressure \cite{Kamada_NiSSe} and La$_{1-x}$Sr$_x$VO$_3$ \cite{Miyasaka_LSVO}, for the metallic compositions near a metal-insulator transition. In these materials, the origin of $T^{1.5}$ dependence of resistivity has been attributed to the existence of strong AF spin fluctuation near Mott transitions. \cite{Moriya_Theory, Julian_Theory} Concerning LNO, the ground state is a correlation-enhanced paramagnetic metal, lying very close to a transition to an AF insulator. Using a combination of electrical resistivity and {\it in situ} PES, the present study indicates that the strong correlations and AF fluctuations result in the temperature-dependent PES of LNO. It is noted that temperature-dependent PES of other nickelates in the $R$NO series, which actually undergo metal-insulator transitions have reported a gradual suppression of spectral weight near $E_{\rm F}$ on decreasing temperature, but a clear gap formation in the valence band DOS has not been observed. \cite{Vobornik_RNO} 

In conclusion, we have performed an {\it in situ} PES measurement on high-quality single-crystal LNO thin films in order to investigate the intrinsic electronic structure of LNO. We find a sharp feature in the PES spectrum, which crosses $E_{\rm F}$. The $E_{\rm F}$ crossing feature shows significant enhancement of spectral weight with decreasing temperature. From a detailed analysis of the temperature dependence of resistivity, the observed spectral changes are attributed to AF fluctuations.

\end{document}